\documentclass[sigconf,nonacm,10pt]{acmart}
\usepackage[english]{babel}
\usepackage{booktabs}
\usepackage{subcaption}
\usepackage{tikz}
\usepackage{xspace}
\usepackage{enumitem}
\usetikzlibrary{arrows.meta,positioning,fit,shapes.geometric}

\acmDOI{}

\acmISBN{}

\author{Jiaying Meng}
\affiliation{%
  \institution{Unaffiliated}
  \country{}
}

\author{Bojie Li}
\affiliation{%
  \institution{Pine AI}
  \country{}
}

\renewcommand{\shortauthors}{Meng and Li}

\newcommand{\sema}{\textsc{Sema}\xspace}

\begin{document}

\title{\sema: Semantic Transport for Real-Time Multimodal Agents}

\begin{abstract}
Real-time multimodal agents transport raw audio and screenshots using networking stacks designed for human receivers, which optimize for perceptual fidelity and smooth playout.
Yet agent models act as event-driven processors with no inherent sense of physical time, consuming task-relevant semantics rather than reconstructing signals in real time. 
This fundamental difference shifts the transport goal from the technical problem of signal fidelity (Shannon--Weaver Level~A) to the semantic problem of meaning preservation (Level~B). 
This mismatch imposes significant overhead. 
In visual pipelines, screenshot upload accounts for over 60\% of end-to-end action latency on constrained uplinks, and in voice pipelines, conventional transport carries massive redundancy, sending 43--64$\times$ more data than needed to maintain task accuracy. 
We present \sema, a semantic transport system that combines discrete audio tokenizers with a hybrid screen representation (lossless accessibility-tree or OCR text, plus compact visual tokens) and bursty token delivery that eliminates jitter buffers.
In simulations under emulated WAN conditions, \sema reduces uplink bandwidth by 64$\times$ for audio and
130--210$\times$ for screenshots while preserving task accuracy within 0.7~percentage points of the raw baseline.
\end{abstract}

\maketitle

\section{Introduction}
\label{sec:intro}


AI agents are increasingly multimodal, maintaining persistent connections that transport rich media streams to the cloud. 
Whether capturing continuous audio for voice interaction (ChatGPT~\cite{chatgpt-voice}, Doubao~\cite{doubao}, Limitless~\cite{limitless}, Plaud~\cite{plaud}, Looki~\cite{looki}) or streaming screenshots for computer use (Atlas~\cite{atlas}, Comet~\cite{comet}, Doubao Phone Assistant~\cite{doubao-assistant}, OpenClaw~\cite{openclaw}), these agents rely on networking stacks designed to transport raw audio and images between the user's device and LLM servers on the cloud.  

These networking stacks inherit two core assumptions from human-facing real-time communication (RTC) systems.  
First, assuming \emph{high sensitivity to signal distortion}, these stacks employ perceptual codecs like Opus~\cite{opus} for audio and WebP for images to optimize for high-fidelity reconstruction. 
Second, assuming \emph{intolerance of timing irregularities}, they mandate jitter buffers and playout schedulers to ensure continuous media playout by smoothing network variations. 

However, neither of these assumptions hold when \emph{the receiver is an agent model (the AI agent's backend LLM) rather than a human}.  
Two intrinsic properties of agent models create an opportunity to rethink media transport: 
\begin{itemize}[leftmargin=*]
\item \textbf{Semantic Requirements.}
Unlike human senses that require rich perceptual signals, agent models effectively process compact, discrete semantic information represented as tokens. 
Recent work demonstrates this by using such discrete representations internally: for audio, end-to-end speech models like CosyVoice~\cite{cosyvoice2} and Qwen3-Omni~\cite{qwen3omni} use audio tokens; 
for vision, specialized visual tokenizers such as FlexTok~\cite{flextok} and structured screen representations via accessibility trees~\cite{computeruseprotocol}. 
These token representations act as information bottlenecks~\cite{tishby1999}, retaining only task-relevant semantics while discarding perceptual redundancy. 
This creates an opportunity to move the generation of these representations to the client, transforming them from a model-internal optimization into a transport-layer optimization that significantly reduces uplink payloads. 

\item \textbf{Event-time Tolerance.}
Unlike human perception that demands continuous playout, agent components process data as ordered event sequences, independent of wall-clock playback timing. 
This decoupling applies across modalities. 
For vision, VLMs process screenshots as discrete events at action steps. 
For audio, on the user-to-agent path, LLMs process tokens in bursts without requiring isochronous audio streams; on the agent-to-user path, TTS systems generate speech in discrete batches, naturally buffering playback.  
This tolerance allows us to abandon the strict jitter-buffering mechanisms required for human listening, simplifying the transport stack. 
\end{itemize}

The mismatch between current networking stacks and these intrinsic properties imposes significant overhead. 
In controlled measurements (Sec. \ref{sec:motivation}), we find that screenshot upload alone accounts for over 60\% of end-to-end action latency in computer-use pipelines at 5\,Mbps uplink. 
This inefficiency stems from solving the wrong problem: conventional transport optimizes for faithful signal reproduction (Shannon--Weaver Level A), whereas agent models only require meaning preservation (Level B). 
A shift to semantic transport unlocks massive gains: by replacing perceptual codecs with compact semantic representations, we can reduce uplink bytes by $\sim$130--210$\times$ for screenshots and $\sim$64$\times$ for voice, all while preserving downstream task accuracy.

To realize this semantic transport paradigm in practice, we present \sema, a semantic transport system that exploits both semantic requirements and event-time tolerance. 
On the uplink, \sema replaces raw media with client-side semantic tokenization. 
For vision, it employs a \emph{hybrid screen representation} that combines lossless structured text (via accessibility trees~\cite{computeruseprotocol}) with compact visual tokens~\cite{layton, flextok}, yielding $\sim$3--5\,KB per screenshot.
For audio, it uses a discrete speech tokenizer~\cite{speechtokenizer} to produce 50--75 token IDs/s ($\sim$500--750\,bps). 
On the server-side, these multimodal inputs are reconstructed for the downstream model.
On the downlink, \sema relocates the vocoder from server to client.

In summary, we make three contributions:
\begin{enumerate}[leftmargin=*]
\item 
We characterize the bandwidth and latency overhead of raw-media transport in multimodal agent pipelines, quantifying the gap between perceptual rate-distortion and task-oriented rate-distortion~\cite{shannon1959}, which defines the minimum bit rate needed at a given quality target.

\item 
We design \sema, a transport system built on two principles for model-centric receivers: (a) semantic requirements, employing client-side hybrid tokenizers to minimize uplink payloads; and (b) event-time tolerance, employing bursty token delivery to eliminate jitter buffers.

\item 
We evaluate \sema via WAN simulation, demonstrating 130--210$\times$ uplink reduction for screenshots and 64$\times$ for audio, while maintaining task accuracy within 0.7~percentage points of the raw baseline.
\end{enumerate}

\section{Background and Motivation}
\label{sec:background}

\subsection{The Transport Bottleneck}
\label{sec:motivation} 
Current multimodal agent pipelines rely on networking stacks that transport raw or perceptually compressed media streams. 
We quantify the cost of this design by measuring the uplink overhead for the two primary modalities: visual streams and audio streams. 

\begin{figure}[tp]
\begin{minipage}[t]{0.49\columnwidth}
\centering
\includegraphics[width=\textwidth]{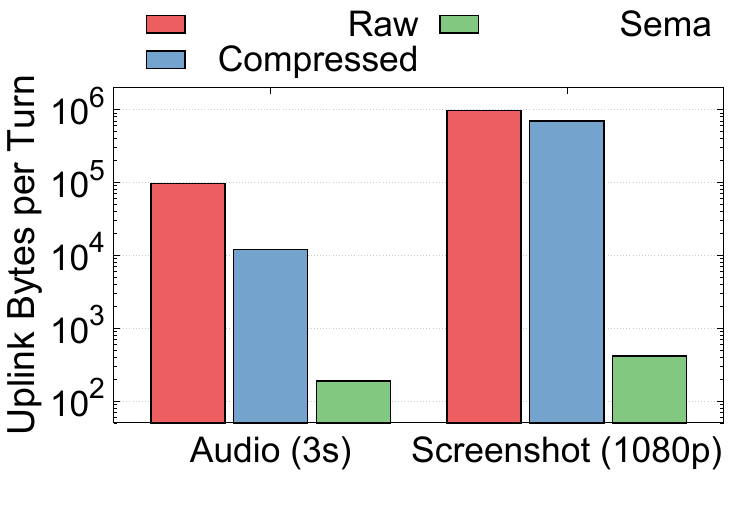}
\caption{Per-turn uplink bytes (log scale).}
\Description{Grouped bar chart on log scale comparing per-turn uplink
  bytes for Raw, Compress, and Sema tokens.}
\label{fig:motivation-bytes}
\end{minipage}\hfill
\begin{minipage}[t]{0.49\columnwidth}
\centering
\includegraphics[width=\textwidth]{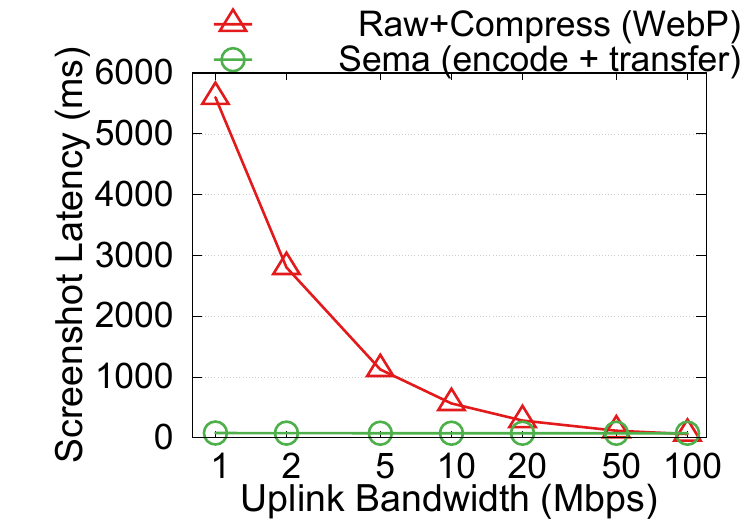}
\caption{Screenshot latency vs.\ uplink bandwidth.}
\Description{Line plot of screenshot latency versus uplink bandwidth.}
\label{fig:motivation-upload}
\end{minipage}
\vspace{-10pt}
\end{figure}

\noindent\textbf{Visual Streams.}
In visual-centric tasks like computer-use, the agent captures a 1080p screenshot at each action step and uploads it as a WebP quality-80 image to a remote VLM. 
Figure~\ref{fig:motivation-bytes} compares per-turn uplink bytes on a log scale: raw PNG requires $\sim$950\,KB and WebP $\sim$700\,KB, whereas \sema's hybrid semantic representation reduces this to $\sim$3--5\,KB, a $\sim$130--210$\times$ reduction.
Figure~\ref{fig:motivation-upload} translates this gap into latency.  
At 5\,Mbps uplink, the WebP upload takes $\sim$1.1\,s, accounting for over 60\% of end-to-end action latency.  
With semantic transport, the combined encode-and-transfer time stays below 100\,ms even at 1\,Mbps, effectively eliminating the upload bottleneck. 

\noindent\textbf{Audio Streams.}
In voice interaction tasks, speech is typically encoded with Opus at 32\,kbps.
In contrast, a semantic layer using SpeechTokenizer~\cite{speechtokenizer} produces just 50 discrete tokens/s ($\sim$62.5\,B/s), achieving a $\sim$64$\times$ reduction (Figure~\ref{fig:motivation-bytes}). 
This massive compression preserves downstream ASR accuracy because the first RVQ layer retains linguistic content~\cite{encodec}, effectively isolating semantics from acoustic details. 

\subsection{Why Semantic Transport Works}
\label{sec:keyideas}

The magnitude of these reductions stems not from more aggressive lossy
compression but from optimizing a fundamentally different distortion
metric: task accuracy rather than signal reconstruction.

\noindent\textbf{Redefining Fidelity: Task Accuracy vs. Perceptual Quality.}
Perceptual codecs (Opus, WebP) minimize distortion as perceived by human senses, but agent models have fundamentally different needs. 
An ASR or multimodal model for voice interaction needs linguistic tokens (phonemes, words, intent), not
waveform quality; a VLM for computer-use tasks needs spatial layout and text content, not
pixel-level texture.
By compressing media into tokenized representations that preserve only what the
downstream model uses, while discarding perceptual details
irrelevant to the agent task, we achieve the 64--210$\times$ bandwidth
ratios shown in
Figures~\ref{fig:motivation-bytes}--\ref{fig:motivation-upload}
without sacrificing task accuracy.
Moreover, because compression efficiency correlates with model
capability~\cite{deletang2023, huang2024compression}, the semantic
capacity of a fixed physical link grows as tokenizer models
improve, which is a scaling property no conventional codec can match.

\noindent\textbf{Decoupling Time: Event Sequences vs. Continuous Playout.}
Human-facing RTC stacks deliver audio continuously at playout rate and
use jitter buffers to smooth timing variation, machinery that exists
solely because human perception is intolerant of discontinuities.
Agent pipelines invert this: AI models are \emph{processing-time
consumers}~\cite{dataflowmodel} that consume ordered token sequences
with no internal clock, so delivery jitter produces no perceptual
artifact at the receiver.
On the user-to-agent path, ASR and LLM decoders recover meaning from
context even when tokens arrive in bursts.  On the agent-to-user
path, TTS models generate speech in discrete batches of several
seconds~\cite{qwen3omni}, creating a natural playout cushion.
Neither direction requires dedicated jitter-buffering machinery~\cite{chatwithai}.
We quantify this tolerance in \S\ref{sec:eval}.

\begin{figure}[t]
  \centering
  \includegraphics[width=0.47\textwidth]{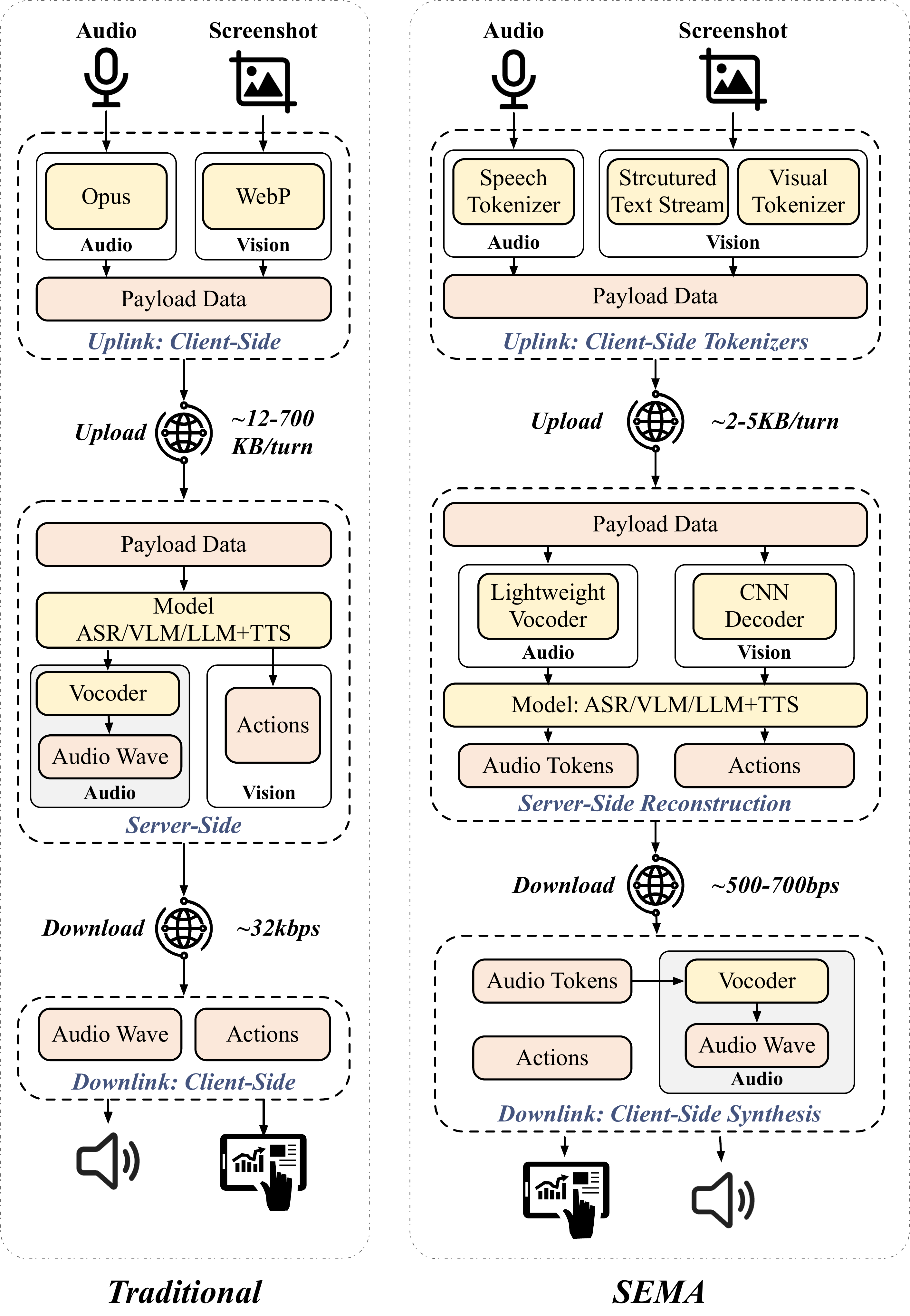}
  \caption{Architectural comparison. (a)~Traditional pipelines send perceptually coded media over the network ($\sim$12--700\,KB per turn) on both paths. (b)~\sema tokenizes on the client (\S\ref{sec:client-tok}), reconstructs on the server (\S\ref{sec:server-recon}), and decodes speech tokens via a client-side vocoder on the downlink (\S\ref{sec:downlink}).  
  Both directions share a lightweight token framing protocol.}
  \label{fig:architecture}
\end{figure}

\section{\sema Design}
\label{sec:design}

\noindent\textbf{Why tokenize, not encode?}
A common misconception is that multimodal large language models
consume all modalities as discrete tokens.  In practice, this is true
only for text.  For images, models such as
Qwen2.5-VL~\cite{qwen25vl} and LLaVA~\cite{llava} pass raw pixels
through a Vision Transformer (ViT) encoder that produces a sequence
of continuous embeddings, one per image patch; for audio, models such
as Whisper~\cite{whisper} use a mel-spectrogram encoder.
These encoders are \emph{not} tokenizers: their output is a
high-dimensional continuous embedding, not a compact codebook index.
Critically, the embedding is typically \emph{larger} than the raw
input, for example, a ViT-L encoder maps a 1024$\times$1024 image
($\sim$3\,MB raw) to 1024 patch embeddings of dimension 1024
($\sim$4\,MB in fp32), and a Whisper encoder maps 3\,s of 16\,kHz
audio ($\sim$96\,KB PCM) to a 1500$\times$1280 embedding
($\sim$7.3\,MB in fp32).
Transmitting the model's internal embeddings would therefore
\emph{increase} network cost, not decrease it.
\sema instead interposes a discrete tokenizer on the client and a
matching reconstruction engine on the server: the tokenizer maps media
to a handful of compact codebook indices (a few hundred bytes per
1024\,px tile, $\sim$62\,B/s for speech), which are transmitted and
then decoded back into a form the model's native encoder can accept.
This design keeps the model's own encoder intact while shrinking the
network payload by orders of magnitude.

Figure~\ref{fig:architecture} shows the \sema architecture.
On the \emph{uplink} (user to agent), client-side tokenizers
(\S\ref{sec:client-tok}) compress audio and screenshots into compact
token sequences.
The server-side reconstructor (\S\ref{sec:server-recon}) converts
these tokens back into the format each downstream model expects.
On the \emph{downlink} (agent to user, \S\ref{sec:downlink}), the
model produces speech tokens that a lightweight client-side vocoder
decodes into audio.
Both directions share a single lightweight framing protocol: each
frame carries a header (modality tag, codebook ID, token count,
sequence number, timestamp) followed by bit-packed codebook indices or
CUP-style~\cite{computeruseprotocol} compact text, and the receiver
demultiplexes by modality tag.

\noindent\textbf{Model Compatibility.}
Voice pipelines span cascaded (ASR, LLM, TTS), end-to-end
speech-to-speech~\cite{defossez2024moshi}, and omni-modal
(Qwen3-Omni~\cite{qwen3omni}, Step-Audio~\cite{stepaudio}) designs;
computer-use pipelines send screenshots to a
VLM~\cite{webvoyager, osworld}.  \sema applies at two tiers:
(1)~\emph{semantic transport}, with client-side tokenization and
server-side reconstruction, works with any model that accepts images
and text (all open-weight VLMs, Whisper~\cite{whisper}, and proprietary
APIs);
(2)~\emph{transport optimization}, using compact payloads and batch-mode
delivery without jitter buffers, is model-agnostic and benefits any
pipeline that produces or consumes media in discrete batches.
As tokenizers and downstream models evolve, \sema swaps components without protocol changes: the frame header's codebook ID negotiates the active tokenizer per session.

\subsection{Uplink: Client-Side Tokenization}
\label{sec:client-tok}

\noindent\textbf{Audio.}
The audio tokenizer runs a discrete speech tokenizer
(SpeechTokenizer~\cite{speechtokenizer} or the first RVQ layer of
EnCodec~\cite{encodec}) on the client.  It produces 50--75 discrete
token IDs per second from a codebook of size 1024, yielding a
bandwidth of 500--750\,bps compared to Opus's 32\,kbps, a 43--64$\times$
reduction.  Encoding is lightweight: SpeechTokenizer processes 1\,s of
audio in $\sim$15--60\,ms depending on hardware.
Only the first RVQ layer is used because agent tasks
require linguistic content but not acoustic fidelity.
Modern end-to-end models (CosyVoice~\cite{cosyvoice2},
Step-Audio~\cite{stepaudio}, Qwen3-Omni~\cite{qwen3omni}) use
discrete speech tokens internally and can serve as drop-in
replacements.

\noindent\textbf{Vision (hybrid screen representation).}
For screenshots, neither a visual tokenizer nor a structured text
representation is sufficient alone.  Visual tokenizers
(Layton~\cite{layton}, FlexTok~\cite{flextok}) produce extremely
compact codes, but small labels, button captions, form content, and text details are lost or garbled after
decoding~\cite{chameleon, ocrvqgan}, making text-heavy tasks such as
form filling or reading notifications fail.  Conversely, the
accessibility tree~\cite{computeruseprotocol} captures text and
element identity perfectly but \emph{discards visual layout}, so
tasks that depend on visual context cannot be solved from text alone.

\sema therefore combines the two streams so that each compensates for
the other's weakness.

\emph{Structured text stream.}
The client reads the platform accessibility API (macOS Accessibility,
Windows UI Automation, Linux AT-SPI, or the web
DOM)~\cite{computeruseprotocol} to obtain element types, labels,
coordinates, and states in a compact text format (e.g.,
\texttt{[e2] button "Back" @132,52 32x32 [click]}).  This yields
$\sim$2--5\,KB per screen with zero encoding compute, because the
accessibility tree is maintained by the OS.
When the accessibility tree is unavailable or incomplete
(canvas-rendered UIs, games, remote desktop), \sema falls back to
lightweight on-device OCR (Apple Vision framework, PaddleOCR;
$\sim$20--50\,ms, $\sim$1--3\,KB).

\emph{Visual token stream.}
To capture visual semantics (spatial layout, iconography, colors) beyond the text stream, the client employs a lightweight tokenizer such as Layton~\cite{layton} (256 tokens per 1024\,px tile) or FlexTok~\cite{flextok} (8--128 variable-length tokens), compressing each tile to $\sim$200--500\,B at $\sim$40\,ms per tile on consumer desktop GPU and $\sim$30--150\,ms on mobile via CoreML~\cite{fastvlm}.
Non-square resolutions (e.g., 1080p, 720p) are handled via \emph{tiling}: each screenshot is split into 1024\,px square regions matching the tokenizer's native input, following the patch strategy used by production vision APIs; a 1080p screenshot yields 2 tiles ($\sim$800\,B total) encoded in a single batched pass, and the structured-text stream carries coordinates in the original pixel space so tile boundaries do not affect grounding.

\emph{Combined payload.}
By fusing these streams, \sema reduces the per-screenshot payload to approximately $\sim$3--5\,KB, which is orders of magnitude smaller than a standard $\sim$700\,KB WebP image.
This massive reduction preserves utility because the structured text provides lossless element identity while visual tokens supply necessary spatial context. 
Our evaluation in \S\ref{sec:eval-pareto} confirms this synergy. 
Visual tokens alone yield only 75.5\% accuracy on text-heavy tasks, but adding the text stream restores performance to 93.3\%, within 0.7~pp of the raw baseline (94.0\%).

\subsection{Server-Side Reconstruction}
\label{sec:server-recon} 
The \sema server accepts hybrid token payloads and reconstructs multimodal input before feeding \emph{both} image and text to the downstream model~\cite{setofmarks, omniparser}.
On the \emph{audio path}, the server handles two scenarios: for models requiring waveforms, it employs a lightweight vocoder to reconstruct audio from codebook embeddings ($\sim$5--10\,ms); for native multimodal models, it forwards discrete tokens directly. 
On the \emph{vision path}, the server first decodes visual tokens via a single-pass CNN, which takes about $\sim$30\,ms using Layton~\cite{layton}. 
It then overlays Set-of-Marks annotations derived from the structured text stream. 
This composite input, consisting of an annotated image for spatial context and structured text for precise element identity, is subsequently fed to the VLM. 
Since the structured text stream ensures accurate content identity, the visual stream only needs to convey spatial context, relaxing the requirement for pixel-perfect reconstruction. 
Overall, the entire reconstruction process adds only $\sim$30--35\,ms of server-side latency.

\subsection{Downlink: Client-Side Synthesis}
\label{sec:downlink}

The downlink carries two types of model output: \emph{audio} for
voice pipelines and \emph{action commands} (e.g., click coordinates,
typed text) for computer-use pipelines.
Action commands are already compact text (tens of bytes per action),
so they need no special optimization in either pipeline.

For audio, the key change is relocating the vocoder from server to
client.  In a traditional pipeline, the server's TTS module generates
speech tokens, a server-side vocoder (e.g.,
HiFi-GAN~\cite{hifigan}) converts them to a waveform, and the
waveform is encoded with Opus ($\sim$32\,kbps) for delivery.
\sema skips server-side vocoding and Opus encoding: the TTS module,
either diffusion-based or multi-codebook autoregressive~\cite{qwen3omni}, produces discrete speech tokens in batches of 3--5\,s and sends them directly to the client
($\sim$500--750\,bps).  A lightweight client-side vocoder then
decodes the tokens into audio locally.
Because each batch amounts to only a few hundred bytes delivered in a single burst every few seconds, no dedicated jitter buffer is needed.

\begin{figure*}[tp]
  \begin{minipage}[t]{0.32\textwidth}
  \centering
  \includegraphics[width=\textwidth]{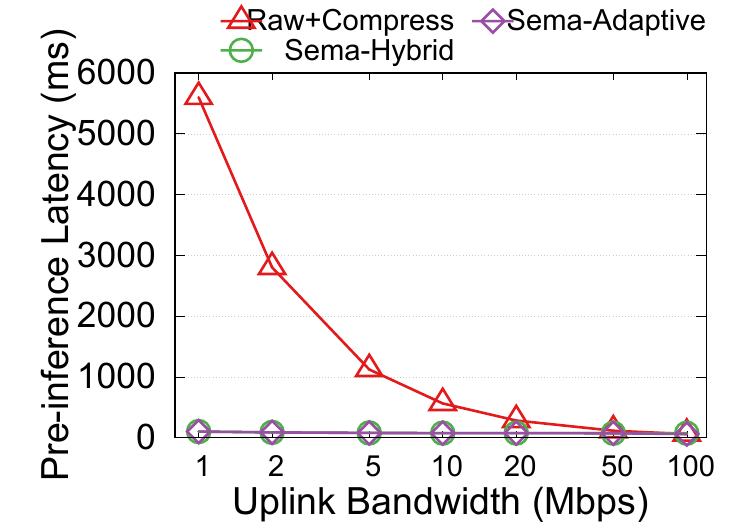}
  \caption{Pre-inference latency (encode + transfer + server decode,
    excluding constant model inference) vs.\ uplink bandwidth
    (RTT\,=\,50\,ms).}
  \Description{Line plot of pre-inference latency versus uplink bandwidth
    from 1 to 100 Mbps.  Raw+Compress decreases steeply from 5.6\,s at
    1\,Mbps; Sema-Hybrid is near-constant at $\sim$70--94\,ms.}
  \label{fig:latency}
  \end{minipage}\hfill
  \begin{minipage}[t]{0.32\textwidth}
  \centering
  \includegraphics[width=\textwidth]{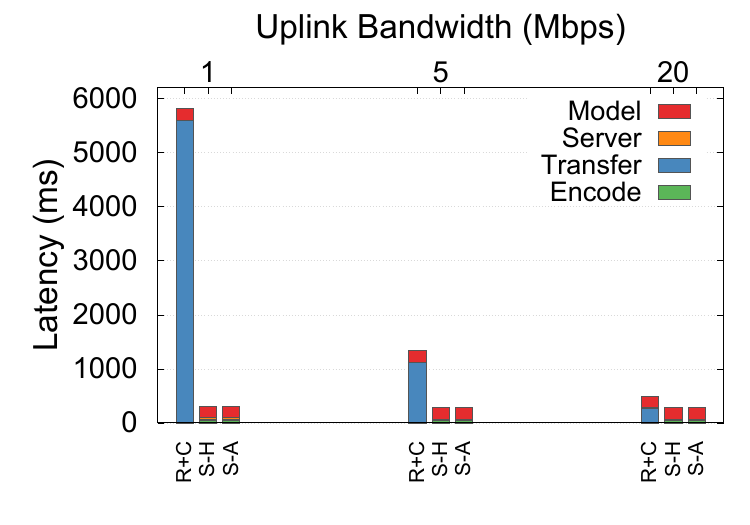}
  \caption{Latency breakdown by pipeline stage at three uplink
    bandwidths.}
  \Description{Stacked bar chart with three bandwidth groups, each with
    bars for Raw+Compress and Sema-Hybrid.
    Transfer dominates Raw+Compress at low bandwidth; client encode +
    server reconstruction is the near-constant cost for Sema.}
  \label{fig:latency-breakdown}
  \end{minipage}\hfill
  \begin{minipage}[t]{0.32\textwidth}
  \centering
  \includegraphics[width=\textwidth]{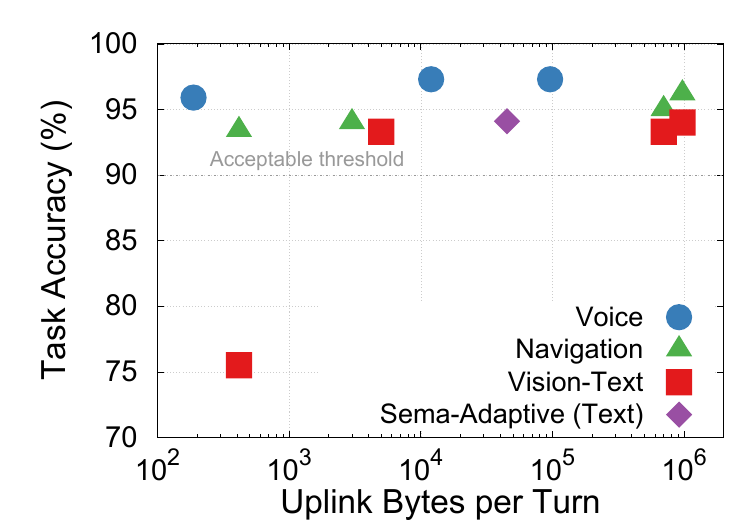}
  \caption{Rate--accuracy tradeoff across methods and modalities
    (log x-axis).}
  \Description{Scatter plot with log x-axis (bytes per turn) and linear
    y-axis (accuracy).  Three legend series: Voice (circles), Navigation
    (triangles), and Vision-Text (squares).
    Within each task series, rightmost points are raw/compressed
    and leftmost are semantic variants.  Vision-text visual-tokens-only
    point drops to 75.5\%; adding structured text recovers to 93.3\%.}
  \label{fig:pareto}
  \end{minipage}
\end{figure*}

\section{Evaluation}
\label{sec:eval}

\subsection{Setup}

We conduct preliminary evaluation of \sema through simulation: each pipeline component
(tokenizer, network transfer, server reconstruction) is modeled using
measured per-component latencies and payload sizes, composed under
emulated network conditions.
An end-to-end implementation is future work.

\noindent\textbf{Tokenizers.}
For audio we use SpeechTokenizer~\cite{speechtokenizer} (first RVQ
layer, 50 tokens/s, codebook 1024); CosyVoice~\cite{cosyvoice2} is a
drop-in alternative at 25\,Hz.  For vision we use
Layton~\cite{layton} (256 tokens per 1024$\times$1024 image with a
1--2 step consistency decoder); FlexTok~\cite{flextok} (variable
8--128 tokens) is cited as an alternative that adapts the token budget
to image complexity.

\noindent\textbf{Models.}
We use open-weight models: Whisper-large-v3~\cite{whisper} for ASR
(via vocoder-reconstructed audio) and
Qwen2.5-VL-7B~\cite{qwen25vl} for computer-use action prediction
(via reconstructed annotated images + structured text).

\noindent\textbf{Workloads and metrics.}
(1)~Voice: 200 turns from LibriSpeech
test-clean~\cite{librispeech}, simulated turn-taking; metric is ASR word-error rate (WER).
(2)~Vision-navigation: 100 web-browsing tasks from the navigation
subset of OSWorld~\cite{osworld}, targeting layout actions (clicking
buttons, navigating menus); metric is exact-match action accuracy against ground-truth click/type events.
(3)~Vision-text: 50 tasks requiring reading small on-screen text or
filling forms, drawn from the productivity subset of OSWorld; metric is task success rate (correct field values submitted).

\noindent\textbf{Baselines.}
(a)~\textbf{Raw}: WebRTC for audio, HTTP upload for PNG screenshots.
(b)~\textbf{Raw+Compress}: Opus at 32\,kbps for audio, WebP quality-80
for 1080p screenshots ($\sim$700\,KB).
(c)~\textbf{Sema-Static}: always uses semantic transport (visual tokens
only, no structured text).
(d)~\textbf{Sema-Hybrid}: structured text (accessibility tree or OCR) +
visual tokens for screenshots, discrete tokens for audio.

\subsection{Bandwidth Reduction}
Table~\ref{tab:bandwidth} reports per-turn uplink bytes.
Discrete audio tokens achieve a 64$\times$ reduction over Opus.
Sema-Static achieves $\sim$840$\times$ over WebP for screenshots (2 tiles at 1080p);
Sema-Hybrid trades some of that ratio (130--210$\times$) for lossless
text fidelity via structured text.
These values report application-layer payload bytes; the shared
token-frame header adds only 17\,B per frame.
\begin{table}[tp]
  \centering
  \caption{Per-turn uplink bytes (median).  Compression ratios are
    relative to Raw+Compress.}
  \label{tab:bandwidth}
  \small
  \setlength{\tabcolsep}{4pt}
  \begin{tabular}{lrrr}
  \toprule
  \textbf{Method} & \textbf{Audio} & \textbf{Screen-} & \textbf{Ratio} \\
                   & \textbf{(3\,s turn)} & \textbf{shot} & \\
  \midrule
  Raw (PCM / PNG)       & 96\,KB  & 950\,KB & \\
  Raw+Compress          & 12\,KB  & 700\,KB & 1$\times$ (ref) \\
  \sema-Static (tokens) & 188\,B  & 832\,B  & 64$\times$ / 841$\times$ \\
  \sema-Hybrid          & 188\,B  & 3--5\,KB & 64$\times$ / 130--210$\times$ \\
  \bottomrule
  \end{tabular}
  \vspace{-5pt}
  \end{table}

\subsection{Latency Improvement}
\label{sec:eval-latency}

\begin{figure*}[tp]
  \begin{minipage}[t]{0.32\textwidth}
  \centering
  \includegraphics[width=\textwidth]{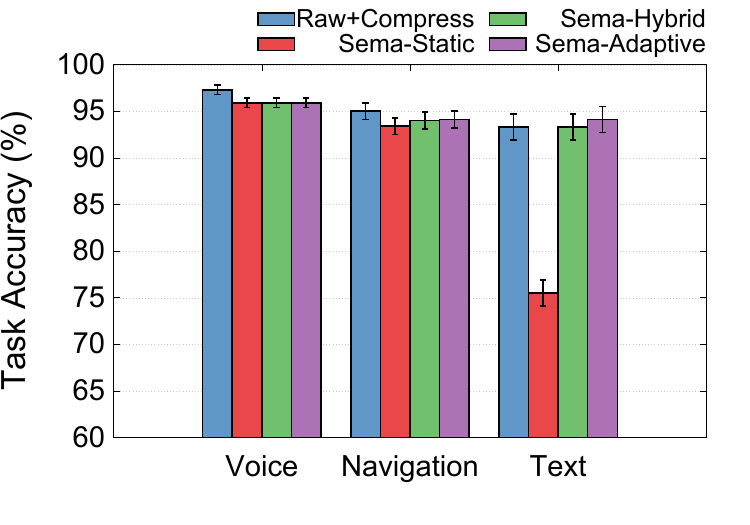}
  \caption{Task accuracy by workload category (95\% CIs).}
  \Description{Grouped bar chart.  Voice and navigation bars are similar
    across methods; Vision-Text shows a large drop for Sema-Static,
    recovered by Sema-Hybrid (lossless structured text).}
  \label{fig:task-accuracy}
  \end{minipage}\hfill
  \begin{minipage}[t]{0.32\textwidth}
  \centering
  \includegraphics[width=\textwidth]{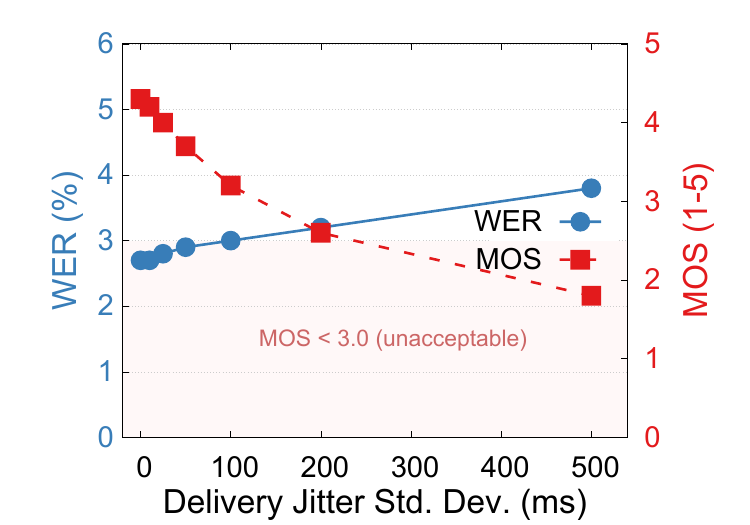}
  \caption{Uplink jitter tolerance: WER (left) vs.\ MOS (right).}
  \Description{Dual-axis line plot.  WER rises gently from 2.7\% to
    3.8\% while MOS falls steeply from 4.3 to 1.8 as jitter increases.}
  \label{fig:jitter-uplink}
  \end{minipage}\hfill
  \begin{minipage}[t]{0.32\textwidth}
  \centering
  \includegraphics[width=\textwidth]{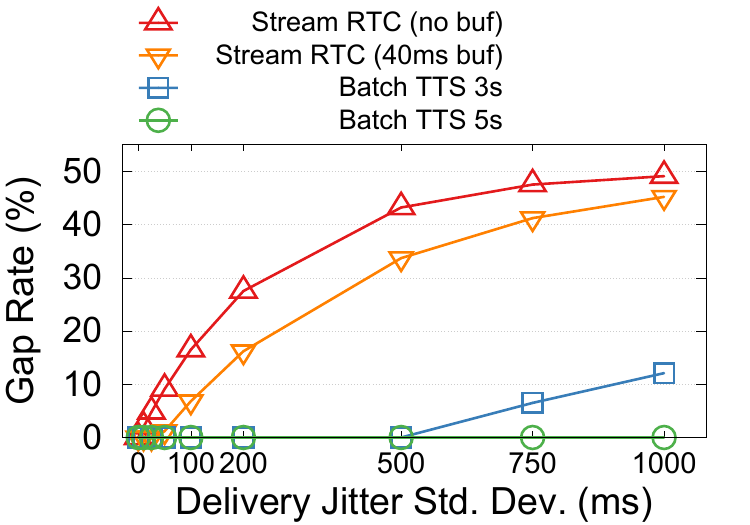}
  \caption{Downlink gap rate vs.\ delivery jitter.}
  \Description{Line plot.  Streaming-RTC gap rate rises steeply; batch
    TTS (3\,s and 5\,s) stays at zero up to 500--1000\,ms jitter.}
  \label{fig:jitter-downlink}
  \end{minipage}
\end{figure*}

Figures~\ref{fig:latency} and~\ref{fig:latency-breakdown} show
end-to-end latency for visual navigation tasks across 1--100\,Mbps uplink.
At 5\,Mbps, Raw+Compress takes $\sim$1.1\,s per 1080p screenshot.
\sema-Hybrid with an accessibility tree totals $\sim$75\,ms
(encode $\sim$40\,ms + transfer $\sim$5\,ms + server decode
$\sim$30\,ms); with OCR fallback, $\sim$105\,ms.
At 1\,Mbps the gap widens: Raw+Compress takes $\sim$5.6\,s while
\sema stays below 100\,ms.
The breakdown (Figure~\ref{fig:latency-breakdown}) confirms that
network transfer dominates Raw+Compress at low bandwidth, while \sema
shifts the bottleneck to a combined $\sim$70--100\,ms encode/decode
cost.  Beyond $\sim$80\,Mbps, raw upload becomes fast enough that the
encoding overhead makes \sema's advantage marginal.

For voice, the latency gap is narrower because audio payloads are
already small.  A 3\,s Opus turn (12\,KB) transfers in $\sim$96\,ms at
1\,Mbps; \sema's 188\,B payload transfers in $<$2\,ms, but adds
$\sim$45--180\,ms client encode and $\sim$8\,ms server vocoder decode,
yielding a net saving only below $\sim$2\,Mbps.  At constrained
uplinks (mobile, IoT), the 64$\times$ bandwidth reduction still
translates to meaningful latency and capacity gains.

\vspace{-0.3cm}
\subsection{Rate--Accuracy Tradeoff}
\label{sec:eval-pareto}
Figure~\ref{fig:pareto} plots uplink bytes per turn against
downstream task accuracy.  The Voice, Navigation, and Vision-Text
series each contain raw, compressed, and semantic transport variants.
For voice and navigation, \sema points fall
in the Pareto-optimal region: near-baseline accuracy at
orders-of-magnitude lower bandwidth.  For vision-text, the
visual-tokens-only point ($\sim$832\,B) drops to 75.5\%, but adding
lossless structured text ($\sim$5\,KB) recovers accuracy to
93.3\%, within 0.7~pp of the 94.0\% raw baseline.

Figure~\ref{fig:task-accuracy} breaks down accuracy by category.
On voice, SpeechTokenizer increases WER from 2.7\% to 4.1\%, a
modest degradation acceptable for conversational agents.
On navigation, action accuracy is within 2~pp of raw.  Sema-Static
drops 12--18~pp on text-heavy tasks~\cite{chameleon, ocrvqgan};
Sema-Hybrid recovers near-baseline accuracy via lossless structured
text.

\subsection{Event-Time Tolerance}
\label{sec:eval-jitter}

Section~\ref{sec:keyideas} argues that agent models consume ordered
event sequences with no internal clock, so delivery jitter should not
degrade task performance.
For vision this holds trivially: VLMs process each screenshot as a
discrete event at each action step, so inter-frame timing variation
has no effect on model input.
Audio is the more demanding case because it is inherently temporal;
we validate both directions below.

\noindent\textbf{Uplink (user to agent).}
Figure~\ref{fig:jitter-uplink} compares the effect of delivery jitter
on two receivers: an ASR model (WER) and a human listener (MOS via
PESQ~\cite{pesq}).
At 200\,ms jitter, MOS drops from 4.3 to 2.6 (40\% degradation)
while WER increases only from 2.7\% to 3.2\% (18\% relative).
At 500\,ms jitter, MOS collapses to 1.8 (below acceptable quality)
while WER remains at 3.8\% (still usable).
The gap confirms that ASR models recover meaning from linguistic
context despite signal-level artifacts.

\noindent\textbf{Downlink (agent to user).}
Figure~\ref{fig:jitter-downlink} measures the playout gap rate as a
function of inter-batch delivery jitter.
With 3\,s TTS batches, playout remains gap-free up to $\sim$500\,ms of
jitter; with 5\,s batches, up to $\sim$1000\,ms.  By contrast, a
traditional RTC system exhibits gaps
under as little as 50\,ms of jitter when jitter buffers are removed.

\vspace{-0.2cm}
\section{Discussion}
\label{sec:discussion}

\noindent\textbf{Text fidelity and client encoding cost.}
The hybrid representation eliminates the primary weakness of pure
visual tokenization by transmitting text losslessly via the
accessibility tree or OCR.
DeepSeek-OCR~\cite{deepseekocr} (97\% accuracy at 10$\times$ compression) and
OCR-VQGAN~\cite{ocrvqgan} continue to close the remaining gap for
pure visual encoders.
All client-side encoding costs (\S\ref{sec:design}) are sub-150\,ms,
keeping the total encode-plus-transfer budget well within interactive
latency targets.

\noindent\textbf{Model compatibility.}
\sema's reconstruct-then-feed approach is model-agnostic: the server
produces a standard annotated image and structured text, compatible
with any VLM (open-weight or proprietary API).  For audio, models
that natively accept discrete tokens can bypass reconstruction
entirely.

\noindent\textbf{Scaling with model progress.}
Conventional codecs are static engineering artifacts whose efficiency
is bounded by fixed source-coding limits.  Semantic transport inherits
ML scaling laws: as tokenizers improve, the transport layer compresses
further automatically.  FlexTok~\cite{flextok} already achieves
FID~$<$~2 with 8--128 tokens; Layton~\cite{layton} reconstructs
1024\,px images from 256 tokens; structured screen
parsers~(CUP~\cite{computeruseprotocol}) continue to advance.
Each gain in model-side representation directly reduces network cost.

\noindent\textbf{From model compression to transport compression.}
The ML community has built increasingly compact representations of
multimodal data, including accessibility trees~\cite{computeruseprotocol},
discrete speech tokens~\cite{cosyvoice2, stepaudio}, and compact visual
tokens~\cite{flextok, layton}.  
\sema's contribution is not a new tokenizer; it is relocating existing model-side compression to the
client and designing the systems infrastructure 
to make this practical for real-time Internet
transport.

\noindent\textbf{Future work.}
Our evaluation is simulation-based; an end-to-end prototype is needed to validate compute budgets on heterogeneous clients, tail latency under real networks, and loss resilience beyond the structured-text fallback (FEC across RVQ layers, selective retransmission remain open).
Baselines use standard WebP/Opus; RoI-WebP or HEVC screen-content extensions could narrow the margin but still target pixel-level distortion, leaving the orders-of-magnitude gap intact.

\vspace{-0.3cm}
\section{Related Work}
\label{sec:related}

\noindent\textbf{AI-oriented RTC.}
Wu et al.~\cite{chatwithai} share our vision, proposing \emph{Context-Aware Video Streaming} (CLIP-guided H.265 RoI, no jitter buffers), but dismiss client-side token streaming as infeasible, arguing MLLMs require continuous embeddings and discrete tokens incur prohibitive quantization loss.
\sema is precisely that demonstration: a \emph{reconstruct-then-feed} server sends discrete tokens on the wire yet decodes them back to continuous pixels before the VLM's native ViT, leaving the model interface untouched; residual loss is absorbed by a lossless accessibility-tree stream which is critical for text understanding. \sema reaches 93.3\% vs.\ 94.0\% raw on the hardest text-heavy workload (\S\ref{sec:eval-pareto}).

\noindent\textbf{Semantic communication.}
DeepSC~\cite{deepsc} introduced deep semantic coding for text;
subsequent work extended the principle to dynamic visual
data~\cite{deepsc-dynamic} and developed coding-theoretic
formulations~\cite{theory-semcom}.  Token
Communications~\cite{tokcom} and VLF-MSC~\cite{vlfmsc} apply
multimodal tokens to the semantic communication setting.
This literature primarily targets wireless links and joint
source-channel coding; \sema applies the same principle to Internet
transport for agent pipelines: transmit meaning, not pixels or waveforms.

\noindent\textbf{Discrete tokenizers.}
Audio RVQ tokenizers (SoundStream~\cite{soundstream},
EnCodec~\cite{encodec}, SpeechTokenizer~\cite{speechtokenizer})
compress speech into discrete codebook indices, enabling \sema to transmit only the first (semantic) layer.
End-to-end models
(CosyVoice~\cite{cosyvoice2}, Step-Audio~\cite{stepaudio},
Qwen3-Omni~\cite{qwen3omni}) build on this foundation.
Visual tokenization descends from VQ-VAE~\cite{vqvae} and
VQGAN~\cite{vqgan}; recent work (FlexTok~\cite{flextok},
Layton~\cite{layton}, OCR-VQGAN~\cite{ocrvqgan}) dramatically
improves fidelity at low token counts.
Conventional codecs (HEVC~\cite{hevc}, screen-content
extensions~\cite{hevcscc}, Salsify~\cite{salsify},
learned compression~\cite{balle2018hyperprior}) all optimize for
pixel-level reconstruction.  \sema shows that discrete tokenizers
developed for model efficiency also serve as transport-layer
compressors when computed on the client.

\noindent\textbf{Agent workloads and screen representations.}
WebVoyager~\cite{webvoyager} and OSWorld~\cite{osworld} benchmark
multimodal agents on GUIs.
OmniParser~\cite{omniparser}, Set-of-Marks~\cite{setofmarks},
GUI-Actor~\cite{guiactor}, and AgentOCR~\cite{agentocr} develop
compact screen representations that reduce model context and improve
grounding.  CUP~\cite{computeruseprotocol} standardizes
accessibility-tree encoding across platforms.
\sema extends these techniques from a model-efficiency role to a
transport-layer role by computing them on the client.

\vspace{-0.3cm}
\section{Conclusion}
\label{sec:conclusion}

When the receiver is a model rather than a human, two transport
axioms change: perceptual fidelity gives way to task-relevant meaning, and wall-clock continuous delivery gives way to event-time bursty delivery.
\sema acts on both: it relocates semantic tokenization to the client
and delivers tokens in bursts rather than streams.
In simulations, \sema reduces uplink bandwidth by 64$\times$ for
audio and 130--210$\times$ for screenshots while maintaining task
accuracy within 0.7~percentage points of the raw baseline.

\bibliographystyle{ACM-Reference-Format}
\bibliography{reference}

\end{document}